\documentclass[aps,prl,twocolumn,epsfig]{revtex4}
\usepackage{amssymb}
\usepackage{amsmath}
\usepackage[dvips]{graphicx}
\usepackage{dcolumn}
\usepackage{bm}

\input{tcilatex}

\begin{document}

\ \ 

\medskip
\title{Inhomogeneous vortex matter}
\author{J.R. Anglin$^{1}$ and M. Crescimanno$^{2}$ }

\affiliation{$^{1}$M.I.T.-Harvard Center for Ultracold Atoms, 77 Massachusetts
Ave., Cambridge MA 02139,\\ 
$^{2}$Center for Photon Induced Processes, 
Physics Department, Youngstown State University, Youngstown, OH 12345.}

\date{\today}

\begin{abstract}
We present a generalization of the continuum theory of vortex matter
for non-uniform superfluid density. This theory explains the striking
regularity of vortex lattices observed in Bose-Einstein condensates, and
predicts the frequencies of long-wavelength lattice excitations.
\end{abstract}


\maketitle


Dense lattices of quantized vortices in rotating Bose-Einstein condensates
(BECs) \cite{MITSci, JILAept} are strikingly more regular than finite vortex
arrays in homogeneous superfluid \cite{CampbellZiff} (see Fig. 1), even
though BEC densities vary greatly over the sample. This Letter generalizes the
Feynman-Tkachenko \cite{Feynman, T1} continuum theory of `vortex matter' to
cases in which the condensate density varies slowly on the scale of the
lattice spacing. This theory explains 
the lattices' surprising regularity, and find
pronounced effects of nonuniform density on lattice excitations. 
\begin{figure}[h]
\includegraphics[width=0.95\linewidth]{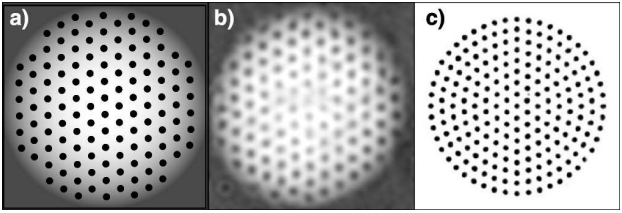}
\caption{(a) Static lattice according to Eq.(\ref{static}), translated and
rotated to match (b) experimental data courtesy of J.R. Abo-Shaeer. Compare
(c) vortex array in constant $\protect\rho $ ($217_{2}$ from Fig. 5 of Ref. 
\protect\cite{CampbellZiff}).}
\end{figure}

We consider a two-dimensional regular array of vortices, whether realized in
a very oblate BEC, or as parallel vortex lines in a prolate one. Denoting
the lattice length scale by $b$ and introducing dimensionless complex
co-ordinates $bz=x+iy$, a regular lattice has the positions $z_{jk}$\ of
parallel vortex lines given by $z_{jk}=z_{jk}^{0}\equiv k+\tau j$ for $\tau
=\tau _{1}+i\tau _{2}$ ($\tau _{2}>0$ and $\tau _{i}$ real). Much is known
about vortex lattices in superfluids of constant density $\rho $, simply
from incompressible hydrodynamics; and it is all simplified by using
dimensionless variables, expressing time, velocity, and energy in lattice
units $Mb^{2}/\hbar $, $\hbar /(Mb)$, $\hbar ^{2}/(Mb^{2})$ respectively,
for $M$ the mass of the particles composing the superfluid. The regular
triangular case $\tau =(1+i\sqrt{3})/2$ is the ground state \cite{T1} of a
sample rotating at dimensionless rate $\Omega =\pi /\tau _{2}$. The
irrotational velocity field $v\equiv v_{x}+iv_{y}$ consists of a fine field $%
v_{f}$, which is periodic on the lattice scale, plus a coarse field $v_{c}$
obeying Feynman's criterion \cite{Feynman} 
\begin{equation}
\partial _{z}v_{c}-\partial _{\bar{z}}\bar{v}_{c}=2\pi i\rho _{V}
\label{Feyn1}
\end{equation}
where for a regular lattice the vortex density $\rho _{V}$ is $1/\tau _{2}$.
\ (In our complex notation 2$\partial _{z}A=b(\vec{\nabla}\cdot \vec{A}+i%
\vec{\nabla}\times \vec{A})$ for any $A$. \ We assume counter-clockwise
rotation.) \ And long-wavelength excitations of the lattice, $%
z_{jk}(t)\rightarrow z_{jk}^{0}+D\left( z_{jk}^{0},\bar{z}_{jk}^{0};t\right) 
$ obey the wave equations \cite{T1} 
\begin{eqnarray}
i\partial _{t}\left( \partial _{z}D-\partial _{\bar{z}}\bar{D}\right)
&=&2\Omega \left( \partial _{z}D+\partial _{\bar{z}}\bar{D}\right)
\label{TT1} \\
i\partial _{t}\left( \partial _{z}D+\partial _{\bar{z}}\bar{D}\right) &=&-%
\frac{1}{2}\partial _{\bar{z}z}\left( \partial _{z}D-\partial _{\bar{z}}\bar{%
D}\right) .  \label{TT2}
\end{eqnarray}
(Extensions to three-dimensional vortex matter \cite{Baym, Fetter} become
considerably more complicated.)

Although current dilute gaseous BECs are compressible fluids governed by the
Gross-Pitaevskii equation (GPE) \cite{gp}, much of BEC vortex physics can be
cast into a simpler hydrodynamic form. In our dimensionless variables, the
GPE in a frame rotating about the co-ordinate origin may be written 
\begin{eqnarray}
\partial _{t}\rho &=&\Omega \partial _{\phi }\rho -\left[ \partial
_{z}\left( \rho v\right) +\partial _{\bar{z}}\left( \rho \bar{v}\right) %
\right]  \label{rhodot} \\
\partial _{t}v &=&\partial _{\bar{z}}\left[ |v-i\Omega z|^{2}+2V+2g\rho -%
\frac{\partial _{\bar{z}z}\sqrt{\rho }}{\sqrt{\rho }}\right]  \label{vdot} \\
V &\equiv &V_{trap}-\frac{\Omega ^{2}}{2}|z|^{2}  \notag
\end{eqnarray}
where $v$ is still the lab-frame velocity, $\phi $ is the polar angular
co-ordinate, $V_{trap}\left( z,\bar{z}\right) $ is the trap potential, and $%
g $ is the dimensionless 2D coupling, determined by atomic and trap
parameters. Except very near vortex cores, the rotating-frame velocity $%
v-i\Omega z$ is of order unity; and $\rho ^{-1/2}\partial _{\bar{z}z}\sqrt{%
\rho }$ is no larger, except in cores and sample edges, which may be treated
separately as boundary layers. In current experiments the\textit{\ healing
length} $\xi \equiv \left( g\rho \right) ^{-1/2},$ which sets the vortex
core size, is everywhere else much smaller than $b$.

So, outside vortex cores, the leading order results in the co-rotating frame
are 
\begin{eqnarray}
\rho &=&\text{const.}-\frac{V}{g}  \label{TF} \\
\Omega \partial _{\phi }\rho &=&\partial _{z}\left( \rho v\right) +\partial
_{\bar{z}}\left( \rho \bar{v}\right) .  \label{cont}
\end{eqnarray}
Thus $\rho $ depends on $v$ only through the centrifugal modification of the
trap potential. The velocity field, $v$ is determined by (\ref{cont}), plus
the condition of irrotationality except at the quantized vortex cores: 
\begin{equation}
\partial _{z}v-\partial _{\bar{z}}\bar{v}=2\pi i\sum_{jk}\delta ^{2}\left(
z-z_{jk}\right) .  \label{curl}
\end{equation}
This, with $\vec{\nabla}\cdot \vec{v}=0$ instead of (\ref{cont}), is the
starting point for Ref. \cite{T1}. So for constant $V$, and hence uniform $%
\rho $, the results of the incompressible case also apply to BECs.

If $\rho $ varies slowly in space, will not inhomogeneous effects be small?
Not obviously: like $\rho $ itself, the lattice shape might vary slowly, but
change greatly over the whole sample. Indeed, even for constant $\rho $ in
the ground state of a finite vortex array, Campbell and Ziff \cite
{CampbellZiff} found gradual but cumulatively large distortions. But there
is a basic problem in extending their analysis to non-uniform $\rho $.

Investigations of homogeneous vortex matter have generally relied on the
exact single-vortex solution $v=v_{1}(\bar{z}-\bar{z}_{jk})$ to (\ref{cont})
and (\ref{curl}) for constant $\rho $, which one may simply sum over the
vortex labels $j,k$, because (\ref{cont}) and (\ref{curl}) are linear in $v$%
. For non-constant $\rho $ the familiar $v_{1}(z)=i/\bar{z}$ satisfies (\ref
{curl}) but not (\ref{cont}), and so we do not have the exact single-vortex
solution. Perturbative approximations about $v_{1}(z)=i/\bar{z}$ as an
ansatz break down at distances from the core beyond the length scale of the
density variation \cite{JRA}. So the few-vortex problem in inhomogeneous
BECs becomes analytically intractable. For sufficiently dense lattices,
however, inhomogeneous vortex matter yields to a different approach.

By a `dense' vortex lattice, we mean $\rho =\rho \left( \varepsilon
z,\varepsilon \bar{z}\right) $ for small $\varepsilon $. (For a round
harmonic trap with Thomas-Fermi radius $R$, $\varepsilon =b/R$, giving $%
\varepsilon \sim 0.1$ in current experiments.) We can therefore perturb in $%
\varepsilon $; but to distinguish smallness from slowness, we must use
multiple scale analysis (MSA) \cite{MSA}. This formalism produces
coarse-scale equations of motion for $D$, from which all lattice-scale
physics has been eliminated, in the same sense that high-frequencies are
eliminated by adiabatic methods. These will be our generalizations of (\ref
{cont}) and (\ref{curl}).

The application of MSA leads to a rather involved derivation the details of
which will be reported elsewhere; here we outline its steps and report its
conclusions. We begin by satisfying (\ref{cont}) identically by defining 
\begin{equation}
v=i\Omega z+\frac{i}{\rho }\partial _{\bar{z}}\left( \rho F\right)
\label{Fdef}
\end{equation}
for real $F$. We use vortex-centered co-ordinates: 
\begin{equation}
z=z^{\prime }+D(z^{\prime },\bar{z}^{\prime },t),
\end{equation}
so that $z_{jk}^{\prime }=z_{jk}^{0}$ regardless of $D$. We do this so that
in the $z^{\prime }$ co-ordinates we always have a regular lattice, whose
symmetries we can exploit, even though the physical lattice may be distorted
by excitations, or by a static $D$ field induced by inhomogeneous $\rho $.
In the non-Cartesian $z^{\prime }$ co-ordinates, Eqn. (\ref{curl}) becomes 
\begin{multline}
\mathrm{Re}\left( \partial _{z^{\prime }}\frac{\partial _{\bar{z}^{\prime
}}-2\left( \partial _{\bar{z}^{\prime }}D\right) \partial _{z^{\prime }}}{%
\rho }\rho F\right) =\pi \sum_{jk}\delta ^{2}\left( z^{\prime
}-z_{jk}^{0}\right)  \notag \\
-\Omega \left( 1+\partial _{z^{\prime }}D+\partial _{\bar{z}^{\prime }}\bar{D%
}+\left| 
\begin{array}{cc}
\partial _{z^{\prime }}D & \partial _{z^{\prime }}\bar{D} \\ 
\partial _{\bar{z}^{\prime }}D & \partial _{\bar{z}^{\prime }}\bar{D}
\end{array}
\right| \right)  \label{F'}
\end{multline}
This shows explicitly that only gradients in $D$ affect $F$.

MSA then embeds the physical $z^{\prime }$-plane in a fictitious 4-space of
complex co-ordinates $\zeta ,Z$, as the subspace $\left( \zeta ,Z\right)
=\left( z^{\prime },\varepsilon z^{\prime }\right) $. This provides $%
\partial _{z^{\prime }}\rightarrow \partial _{\zeta }+\varepsilon \partial
_{Z}$, etc., and proceeding perturbatively in $\varepsilon $, we are able to
write explicit solutions (in terms of rapidly converging series) for the $%
\zeta $-dependence of $F$ at every order (we need to go to third). As usual
with MSA, the `gauge freedom' in how functions depend explicitly on the two
extra dimensions is used to remove solutions growing secularly with $\zeta $%
, by constraining the purely $Z$-dependent part of $F$ (which we denote by $%
F_{c}$). Once we restrict back to physical two-space by setting $%
Z=\varepsilon z^{\prime },$ $\zeta =z^{\prime }$, and return from
vortex-fixed $z^{\prime }$ to Cartesian $z$, we recognize the constraint on $%
F_{c}$ as just what is needed to maintain Feynman's condition (\ref{Feyn1}),
for $\rho _{V}$ as perturbed by $D$: 
\begin{equation}
\partial _{z}\frac{\partial _{\bar{z}}\left( \rho F_{c}\right) }{\rho }%
+\partial _{\bar{z}}\frac{\partial _{z}\left( \rho F_{c}\right) }{\rho }%
=-2\Omega \left( \partial _{z}D+\partial _{\bar{z}}\bar{D}\right) \;,
\label{Feyn2}
\end{equation}
where we drop determinant terms quadratic in $D$ (because we will have $D$
order $\varepsilon $ or smaller).

Having solved for $F,$ and hence $v,$ in terms of explicit lattice-periodic
functions and $v_{c}$, we know the local fluid velocity near each vortex.
This fixes the instantaneous vortex translational velocity field $\dot{D}$;
but the fixing is not trivial. Since the hydrodynamic approximation to the
GPE breaks down within $|z-z_{jk}|\sim $ $\xi /b$, in these small regions we
must solve the time-dependent GPE using a different perturbation theory,
based on Taylor-expanding $V$ about $z_{jk}$. Matching the hydrodynamic and
core solutions smoothly together (see Refs. \cite{JRA, RP}) finally yields,
to leading order in $\varepsilon $, 
\begin{eqnarray}
i\rho \dot{D}&=&\frac{1}{2}\left[ \partial _{\bar{z}}\left( \rho \partial
_{z}D+\rho \partial _{\bar{z}}\bar{D}\right) -\partial _{z}\left( \rho
\partial _{\bar{z}}D\right) \right] -\partial _{\bar{z}}\left( \rho
F_{c}\right)  \notag \\
&&+\left( \ln \frac{b}{2\pi \xi }+1.17\right) \partial _{\bar{z}}\rho
-0.20\partial _{z}\frac{\left( \partial _{\bar{z}}\rho \right) ^{2}}{\rho }
\label{main}
\end{eqnarray}
which is expressed in the co-rotating frame.

Eqns. (\ref{Feyn2}) and (\ref{main}) are our main results. The numerical
co-efficients in (\ref{main}) include some numerically evaluated
contributions from the nonlinear core regions (compare with \cite{RP}), and
also functions of $\tau $, generally related to $\theta $-functions,
evaluated for the triangular case $\tau =(1+i\sqrt{3})/2$. For general $\tau 
$\ (\textit{i.e.} for lattices other than the regular triangular), (\ref
{main}) would have several additional terms, such as $B(\tau )\partial _{%
\bar{z}\bar{z}}D$, where $B(\tau )$ is another rapidly converging series.
Modular covariance (a general type of lattice symmetry) of the extra
co-efficients like $B(\tau )$ constrains them to vanish when $\tau =\frac{1+i%
\sqrt{3}}{2}$.

MSA implies that the lattice scale $b$ is to the vortex matter equations
much as the healing length $\xi $ is to the hydrodynamic equations that
underly them. Thus Eqns. (\ref{Feyn2}) and (\ref{main}) should be accurate
except at distances of order $b$ or less from of the edge of a vortex array.
But is this claim really compatible with the results of Campbell and Ziff 
\cite{CampbellZiff} for finite vortex arrays in infinite homogeneous
superfluid? Setting $\rho $ constant, to leading order in $\varepsilon ^{2}$
the pair (\ref{Feyn2}) (\ref{main}) reduces to (\ref{TT1}) and (\ref{TT2}),
and setting $\dot{D}\rightarrow \dot{D}_{0}=0$, we find a wealth of
solutions to these fourth order equations: 
\begin{equation}
D_{0}=\sum_{m=1}^{\infty }\left( a_{m}\bar{z}^{m-1}+(m+1)b_{m}\bar{z}%
^{m}z-b_{m}^{\ast }z^{m+1}\right) 
\end{equation}
for arbitrary complex constants $a_{m},b_{m}$. \textit{A priori} it is
unclear what boundary condition $D$ should respect at the array edge;
however if we assume that the edge should be a uniform circle of vortices,
fitting leads to a unique combination of multipolar distortions with $%
m=6,12,18...$. Figure 2 shows that stopping at only $m=12$ gives quite good
agreement with the first excited state for 217 vortices found in Ref. \cite
{CampbellZiff}. (The ground state differs only in the outermost ring.) 
\begin{figure}[h]
\includegraphics[width=0.9\linewidth]{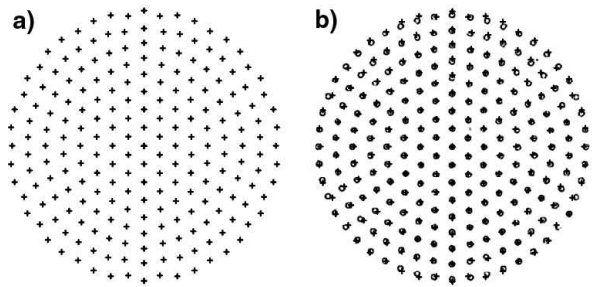}
\caption{Distortions of a finite vortex array in infinite homogeneous
superfluid. Fig. 2a) shows the combination of $m=0,6,$ and 12 solutions that
makes the outer ring most circular and evenly spaced, while 2b) is a)
overlaid on array 217$_{2}$ of Fig. 5 of Ref. \protect\cite{CampbellZiff}.}
\end{figure}

If we set $\rho =\rho _{0}(1-\varepsilon ^{2}|z|^{2})$ for a BEC in a round
harmonic trap, though, the only distortion forced on us by the inhomogeneity
is the very mild 
\begin{equation}
D_{0}={\frac{\sqrt{3}}{4\pi }}{\frac{\varepsilon ^{2}z}{1-\varepsilon
^{2}|z|^{2}}}\left( \ln {\frac{b}{2\pi \xi }}+1.17\right) +\mathcal{O}\left(
\varepsilon ^{4}\right)  \label{static}
\end{equation}
This scarcely visible radial shift of each vortex is shown in Fig. 1, for $%
\varepsilon =1/6=\xi (0)/b$. For vortices very close to the TF surface where
formally $\xi \rightarrow \infty $, (\ref{static}) spuriously predicts large
inward displacements. (Four such vortices have been excised from Fig. 1a.)
Apart from this failure in the lattice-edge boundary layer of thickness $b$,
the accord with experiment is excellent.

Using (\ref{Fdef}) and (\ref{main}), the $v_{c}$ associated with this $%
D_{0} $ is 
\begin{equation}
v_{c0}= i z \left(\Omega - \frac{\varepsilon ^{2}[\ln {\frac{b}{2\pi \xi }}%
+1.17]}{1-\varepsilon ^{2}|z|^{2}}\right).
\end{equation}
This purely azimuthal flow is slightly less than rigid body rotation at $%
\Omega$, and the magnitude of the backflow increases with radius. This
agrees qualitatively with the numerical results shown in Fig. 5 of Ref. \cite
{David}.

Any static distortions forced by boundary conditions, like those of Fig.
(2), would appear as zero-frequency modes among the collective excitations;
so to these we now turn. We write $D=D_{0}+\partial _{\bar{z}}(P+iQ)$,
without loss of generality, for real $P,Q$. Then $\mathrm{Re}\{\partial
_{z}[ $(\ref{main})$/\rho ]\}$ and (\ref{Feyn2}) yield 
\begin{equation}
\partial _{\bar{z}z}\dot{Q}=-2\Omega \partial _{\bar{z}z}P\times \left[ 1+%
\mathcal{O}(\varepsilon ^{2})\right] .
\end{equation}
Since only $D$ is physical, any terms in $P$ annihilated by $\partial _{\bar{%
z}z}$ are of form $f(z)+f^{\ast }(\bar{z})$ and so can be absorbed in $Q$ as 
$i[f^{\ast }(\bar{z})-f(z)]$. \ And since one can easily show that
Laplacian-free terms in $Q$ must be time-independent to leading order in $%
\varepsilon ^{2}$, we can set $P=-\frac{1}{2\Omega }\dot{Q}$.

Introducing polar co-ordinates $re^{i\phi }=\varepsilon z$, so that $\rho
=\rho _{0}(1-r^{2})$, we can write $Q=q_{mn}(r)\cos \left( m\phi -\omega
_{mn}t-\alpha \right) $ (for arbitrary constant $\alpha $ and angular and
radial quantum numbers $m$ and $n$). \ Considering first the case $m=0,$
where $\partial _{\phi }F_{c}=0$, the imaginary part of $e^{-i\phi }\times $(%
\ref{main}) implies 
\begin{equation}
\frac{4\omega _{0n}^{2}}{\varepsilon ^{2}\Omega }\partial _{r}q_{0n}=-\frac{%
\left( \partial _{r}+2r^{-1}\right) }{\left( 1-r^{2}\right) }\left[ \left(
1-r^{2}\right) \left( \partial _{r}-r^{-1}\right) \partial _{r}q_{0n}\right]
\label{m0}
\end{equation}
plus order $\varepsilon .$ Frobenius analysis shows that only one solution
to this second order equation for $\partial _{r}q_{0n}$ is finite at $%
r\rightarrow 0$, and imposing finiteness at $r\rightarrow 1$ as well fixes
the discrete spectrum. \ (Modified behavior within the boundary layer $%
r\gtrsim 1-\varepsilon $ will be able to reconcile our coarse-scale $D$ and $%
v_{c}$ with microscopic boundary conditions, as long as our functions remain
regular; compare the hydrodynamic derivation of collective modes in the
vortex-free BEC \cite{Sandro1}.) Numerical search yields 
\begin{equation}
\omega _{0n}=\varepsilon \Omega \left\{ 0,1.43,2.32,3.18,4.03,...\right\} .
\end{equation}
The zero mode is global rotation of the lattice, $q_{00}=r^{2}$. \ 

These results are unsurprising for Tkachenko waves in a finite cylindrical
system; but for $m\neq 0$, the dynamics is very different. To eliminate $%
F_{c}$ we must take $\mathrm{Im}[\partial _{z}$(\ref{main})$]$, and the
non-constant $\rho $ leaves a first order time derivative on the left side,
implying $\omega _{mn}$ of order $\varepsilon ^{2}$ instead of $\varepsilon
. $ Our leading order equation is thus first order in $t$ but fourth in $r$: 
\begin{eqnarray}
\frac{16m\omega _{mn}}{\varepsilon ^{2}}q_{mn} &=&-\left[ \left( \partial
_{rr}+r^{-1}\partial _{r}-m^{2}r^{-2}\right) -4r\partial _{r}\right]  \notag
\\
&&\times \left( \partial _{rr}+r^{-1}\partial _{r}-m^{2}r^{-2}\right) q_{mn}.
\label{fourho}
\end{eqnarray}
This equation is quite singular (Frobenius analysis shows that of the four
solutions only one is finite at both $r=0,1$); but its differential operator
is Hermitian and self-dual, and its regular solution has a rapidly
converging Frobenius series in $r.$ From (\ref{main}) we see that the radial
component of $v_{c}$ blows up at $r=1$ unless $\left( \partial
_{rr}-r^{-1}\partial _{r}-m^{2}r^{-2}\right) q_{mn}(1)=0$. \ This boundary
condition on the regular solutions $q_{mn}(r)$ fixes the discrete spectrum,
which may be found by summing Frobenius series numerically:

\begin{center}
\begin{tabular}{c|cccccc}
$\frac{\omega _{mn}}{\varepsilon ^{2}\Omega }$ & $m=1$ & 2 & 3 & 4 & 5 & 6
\\ \hline
$n=0$ & 0 & 0.365 & 0.900 & 1.60 & 2.46 & 3.49 \\ 
1 & 2.93 & 5.02 & 7.53 & 10.5 & 13.8 & 17.5 \\ 
2 & 32.0 & 31.3 & 35.6 & 41.6 & 48.8 & 56.8 \\ 
3 & 130. & 105. & 106. & 113. & 124. & 137.
\end{tabular}
\end{center}

\noindent and so on. The translational zero mode is $q_{10}=r$. \ These
eigenvalues of a fourth order equation increase rapidly with radial quantum
number $n$, and as the co-efficients reach order $\varepsilon ^{-1}$ (\ref
{fourho}) becomes invalid, and WKB-like Tkachenko waves will emerge instead.

The only zero-frequency solution satisfying the boundary conditions is the
rigid translation; and so (\ref{static}) is the full static distortion. This
reflects the fact that $\rho $ is so small at the edges of the lattice that
no boundary energies are large enough to influence the bulk lattice.

Only positive $m$ have been reported, because 
$\omega_{-m,n}=-\omega _{mn}$ and replacing $m\rightarrow -m$ leaves our ansatz
for $Q$ unchanged. What this means is that the nonuniform $\rho $ has
drastically split the degeneracy of the two modes that would, for constant $%
\rho $, be proportional to $e^{\pm im\phi }$, with frequency of order $%
\varepsilon $. The linear combination proportional to $\cos (m\phi -\omega
_{mn}t)$ propagates much more slowly, in the co-rotating frame; evidently
the orthogonal combination, which we have not examined, propagates much more
quickly. (The distortion patterns rotate about the origin; the vortices
follow elliptical orbits about their equilibrium positions: see Fig. 3.)
Thus the lowest frequency lattice modes have a first order dynamics, and
only half as many distinct modes as for constant $\rho $.

\begin{figure}[h]
\includegraphics[width=0.9\linewidth]{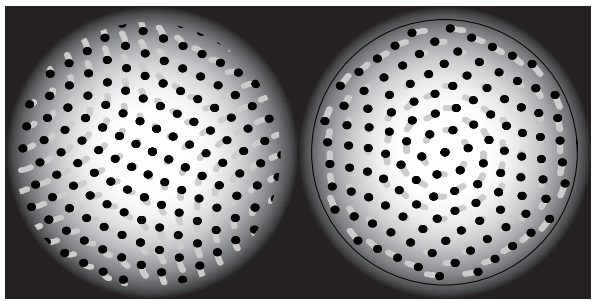}
\caption{Vortex lattice excitations in a round harmonic trap. \ The grey
`trails' indicate vortex motion. \ Left: $m,n=2,0;$ compare Figure 3 b) of
Ref. \protect\cite{JILAept}. Right: $m,n=0,1$, in which motion is almost
purely angular, and much faster.}
\end{figure}

Finally, note that for the quadrupole mode $\omega _{20}\neq 0$ (the zero
eigenvalue solution $\tilde{q}_{20}=r^{2}$ does not satisfy the boundary
condition). \ Since it is this mode which would distort the equilateral
triangular lattice into the moderately different regular lattices that are
also dynamically stable on short wavelengths \cite{T1}, we conclude that
although stable in bulk those lattices are frustrated in the finite system,
and cannot even be stationary without slow but cumulatively large
distortions. \ Reviewing our calculations in this context, it is clear that
the only reason the equilateral triangular lattice does not suffer a similar
fate is the vanishing, due to lattice symmetry, of several awkward terms
from (\ref{main}). Once the threat of cumulative distortion is lifted, it is
not surprising that merely local distortion is of order $\varepsilon ^{2}$.
\ So the regularity of the observed BEC vortex lattices is ultimately due to
their triangular structure. \ An engineer would attribute this to triangular
rigidity, and a mathematician to the fact that the triangular lattice is the 
$Z_{3}$ fixed point of the modular group. \ Physicists are entitled to
arbitrary linear combinations of the two explanations.

This work has been supported by the US National Science Foundation through a
grant for the Center for Ultracold Atoms and through a grant for the
Institute for Theoretical Atomic and Molecular Physics at Harvard University
and the Smithsonian Astrophysical Observatory. MC acknowledges partial
support from the DOE EPSCoR grant DE-FG01-00ER45832, Research Corporation
Cottrell Science Award \#CC5285, and a YSU Research Professorship Award.


\vskip .2in

\begin{thebibliography}{99}
\bibitem{MITSci}  J.R. Abo-Shaeer et al., Science \textbf{292}, 476 (2001).

\bibitem{JILAept}  P. Engels, I. Coddington, P. C. Haljan, and E. A.
Cornell, cond-mat/0204449.

\bibitem{CampbellZiff}  L.J. Campbell and R.M. Ziff, Phys. Rev. \textbf{B20}%
, 1886 (1979).

\bibitem{Feynman}  R.P. Feynman, Ch. 2 in C.J. Gorter, ed., \textit{Progress
in Low Temperature Physics }(North-Holland, Amsterdam, 1955).

\bibitem{T1}  V.K. Tkachenko, Sov. Phys. JETP \textbf{23}, 1049 (1966); 
\textbf{29}, 945 (1969).

\bibitem{Baym}  G. Baym and E. Chandler, J. Low Temp. Phys. \textbf{50}, 57
(1983).

\bibitem{Fetter}  M.R. Williams and A.L. Fetter, Phys. Rev. \textbf{B16},
4846 (1977).

\bibitem{gp}  V. L. Ginzburg and L. P. Pitaevskii, Sov. Phys. JETP \textbf{7}%
, 858 (1958); E. P. Gross, J. Math. Phys. \textbf{4}, 195 (1963).

\bibitem{JRA}  J.R. Anglin, Phys. Rev. \textbf{A65}, 063611 (2002).

\bibitem{MSA}  See, e.g., C.M. Bender and S.A. Orszag, \textit{Advanced
Mathematical Methods for Scientists and Engineers }(McGraw-Hill, New York,
1978), Chapter 11.

\bibitem{RP}  B. Y. Rubinstein and L. M. Pismen, Physica D, \textbf{78}, 1
(1994).

\bibitem{David}  D. L. Feder and C. W. Clark, Phys. Rev. Lett. \textbf{87}
190401 (2001).

\bibitem{Sandro1}  S. Stringari, Phys. Rev. Lett. \textbf{77}, 2360 (1996).
\end{thebibliography}
\end{document}